\documentclass[10pt,draft]{article}
\usepackage{amsfonts,amsmath,amssymb,cite}
\newcommand{\Real}{\mathop{\textrm{Re}}}
\newcommand{\sgn}{\mathop{\textrm{sgn}}}

\begin{document}
\title{Magnetic field-induced electric quadrupole moment \\
in the ground state of the relativistic hydrogen-like atom: \\
Application of the Sturmian expansion of the generalized
Dirac--Coulomb Green function}
\author{Rados{\l}aw Szmytkowski\footnote{Corresponding author. 
Email: radek@mif.pg.gda.pl} \hspace*{0em}
and
Patrycja Stefa{\'n}ska \\*[3ex]
Atomic Physics Division,
Department of Atomic Physics and Luminescence, \\
Faculty of Applied Physics and Mathematics,
Gda{\'n}sk University of Technology, \\
Narutowicza 11/12, 80--233 Gda{\'n}sk, Poland}
\date{\today}
\maketitle
\begin{abstract}
We consider a Dirac one-electron atom placed in a weak, static,
uniform magnetic field. We show that, to the first order in the
strength $B$ of the perturbing field, the only electric multipole
moment induced by the field in the ground state of the atom is the
quadrupole one. Using the Sturmian expansion of the generalized
Dirac--Coulomb Green function [R.~Szmytkowski, J.\ Phys.\ B 30 (1997)
825; erratum 30 (1997) 2747], we derive a closed-form expression for
the induced electric quadrupole moment. The result contains the
generalized hypergeometric function $_{3}F_{2}$ of the unit argument.
Earlier calculations by other authors, based on the non-relativistic
model of the atom, predicted in the low-field region the quadratic
dependence of the induced electric quadrupole moment on $B$.
\end{abstract}
%
%
\section{Introduction}
\label{I}
\setcounter{equation}{0}
In the mid 1950's, Coulson and Stephen \cite{Coul56} pointed out that
the uniform magnetic field should induce the electric quadrupole
moment (EQM) in the hydrogen-like atom. They used the perturbation
theory and found that in the case of the non-relativistic atom, for
field strengths $B$ corresponding to the complete Paschen--Back
effect, the leading term in the expansion of the induced EQM in
powers of the field strength is \emph{quadratic\/} in $B$. For the
atomic ground state, their result, after being translated into the
form conforming both to the definition of the EQM (cf.\ Sec.\
\ref{III} below) and to the notation used in the present work, is
\begin{equation}
\mathcal{Q}_{20}\simeq\mathcal{Q}_{20}^{(2)}
=-\frac{5}{16}\frac{\alpha^{4}ea_{0}^{2}}{Z^{6}}
\frac{B^{2}}{B_{0}^{2}},
\label{1.1}
\end{equation}
where $Ze$ is the nuclear charge, $a_{0}$ is the Bohr radius, $\alpha$
is the Sommerfeld fine-structure constant, whereas
\begin{equation}
B_{0}=\frac{\mu_{0}}{4\pi}\frac{\mu_{\mathrm{B}}}{a_{0}^{3}}
=\frac{\alpha^{2}\hbar}{2ea_{0}^{2}}\simeq6.26\,\mathrm{T}
\label{1.2}
\end{equation}
is the atomic unit of the magnetic induction ($\mu_{0}$ is the vacuum
permeability and $\mu_{\mathrm{B}}$ is the Bohr magneton). Later,
Turbiner \cite[Sec.\ 5]{Turb87} arrived at the same expression for
$\mathcal{Q}_{20}^{(2)}$ for the Schr{\"o}dinger one-electron atom
with $Z=1$ and found also explicitly the next non-vanishing term
(being $\mathcal{Q}_{20}^{(4)}\sim B^{4}$) in the expansion of
$\mathcal{Q}_{20}$ in powers of $B$. Moreover, using the variational
technique, he determined the function $\mathcal{Q}_{20}(B)$ for
magnetic fields ranging from vanishing to superstrong. A decade ago,
Potekhin and Turbiner \cite{Pote01} calculated $\mathcal{Q}_{20}(B)$,
over a still broader range of $B$, by two different methods, one
being the variational approach with a more sophisticated trial
function than the one used in Ref.\ \cite{Turb87}, the other being
based on the expansion of the perturbed electronic wave function in
the Landau orbitals. In the low-field limit, the results of Ref.\
\cite{Pote01} reproduced the quadratic dependence of the induced EQM
on $B$ predicted in Refs.\ \cite{Coul56,Turb87}.

In all the aforementioned works, the atom has been described
non-relativistically. In the present paper, we show that if the
relativity is taken into account and the atomic model adopted is the
one based on the Dirac equation for the electron, it appears that the
weak, static, uniform magnetic field induces in the ground state of
the atom the EQM which, to the lowest order, is \emph{linear\/} in
the perturbing field strength $B$; no other electric multipole
moments are induced in the system to the first order in $B$.
Exploiting the Sturmian expansion of the generalized (or reduced)
Dirac--Coulomb Green function, found by one of us in Ref.\
\cite{Szmy97} and subsequently successfully used in analytical
calculations of various electromagnetic properties of the Dirac
one-electron atom \cite{Szmy02,Szmy04,Miel06,Szmy11,Stef11}, in Sec.\
\ref{IV} we derive a closed-form expression for the induced EQM, in
terms of the generalized hypergeometric function $_{3}F_{2}$ with the
unit argument.
%
%
\section{Preliminaries}
\label{II}
\setcounter{equation}{0}
It has been already stated in the introduction that the system we
shall be concerned with in the present work is the Dirac one-electron
atom. Its nucleus will be assumed to be point-like, infinitely heavy,
spinless, and to carry the electric charge $Ze$. In the absence of
external perturbations, the atomic ground-state energy level
\begin{equation}
E^{(0)}=mc^{2}\gamma_{1},
\label{2.1}
\end{equation}
where
\begin{equation}
\gamma_{\kappa}=\sqrt{\kappa^{2}-(\alpha Z)^{2}},
\label{2.2}
\end{equation}
is two-fold degenerate, the two associated Hamiltonian
eigenfunctions, orthonormal in the sense of
\begin{equation}
\int_{\mathbb{R}^{3}}\mathrm{d}^{3}\boldsymbol{r}\:
\Psi_{\mu}^{(0)\dag}(\boldsymbol{r})\Psi_{\mu'}^{(0)}(\boldsymbol{r})
=\delta_{\mu\mu'},
\label{2.3}
\end{equation}
being
\begin{equation}
\Psi_{\mu}^{(0)}(\boldsymbol{r})
=\frac{1}{r}
\left(
\begin{array}{c}
P^{(0)}(r)\Omega_{-1\mu}(\boldsymbol{n}_{r}) \\
\mathrm{i}Q^{(0)}(r)\Omega_{1\mu}(\boldsymbol{n}_{r}) \\
\end{array}
\right)
\qquad ({\textstyle\mu=\pm\frac{1}{2}}).
\label{2.4}
\end{equation}
The radial functions appearing in Eq.\ (\ref{2.4}), normalized
according to
\begin{equation}
\int_{0}^{\infty}\mathrm{d}r
\left\{[P^{(0)}(r)]^{2}+[Q^{(0)}(r)]^{2}\right\}=1,
\label{2.5}
\end{equation}
are explicitly given by
\begin{subequations}
\begin{equation}
P^{(0)}(r)=-\sqrt{\frac{Z}{a_{0}}\frac{1+\gamma_{1}}
{\Gamma(2\gamma_{1}+1)}}
\left(\frac{2Zr}{a_{0}}\right)^{\gamma_{1}}\mathrm{e}^{-Zr/a_{0}},
\label{2.6a}
\end{equation}
\begin{equation}
Q^{(0)}(r)=\sqrt{\frac{Z}{a_{0}}\frac{1-\gamma_{1}}
{\Gamma(2\gamma_{1}+1)}}
\left(\frac{2Zr}{a_{0}}\right)^{\gamma_{1}}\mathrm{e}^{-Zr/a_{0}},
\label{2.6b}
\end{equation}
\label{2.6}
\end{subequations}
\newline
whereas $\Omega_{\kappa\mu}(\boldsymbol{n}_{r})$, with
$\boldsymbol{n}_{r}=\boldsymbol{r}/r$, are the orthonormal
spherical spinors defined as in Ref.\ \cite{Szmy07}.

In the presence of a weak, static, uniform magnetic field
$\boldsymbol{B}=B\boldsymbol{n}_{z}$, the level $E^{(0)}$ splits into
two. Their energies, to the first order in $\boldsymbol{B}$, are
given by
\begin{equation}
E_{\mu}\simeq E^{(0)}+E_{\mu}^{(1)}
\qquad ({\textstyle\mu=\pm\frac{1}{2}}),
\label{2.7}
\end{equation}
with
\begin{equation}
E_{\mu}^{(1)}=\sgn(\mu)\frac{2\gamma_{1}+1}{3}\mu_{\mathrm{B}}B.
\label{2.8}
\end{equation}
The corresponding wave functions, to the same approximation order,
are
\begin{equation}
\Psi_{\mu}(\boldsymbol{r})\simeq\Psi_{\mu}^{(0)}(\boldsymbol{r})
+\Psi_{\mu}^{(1)}(\boldsymbol{r})
\qquad ({\textstyle\mu=\pm\frac{1}{2}}).
\label{2.9}
\end{equation}
Here, the zeroth-order component $\Psi_{\mu}^{(0)}(\boldsymbol{r})$
is given by Eq.\ (\ref{2.4}) (the space quantization axis is now
chosen to be directed along $\boldsymbol{B}$). The correction 
$\Psi_{\mu}^{(1)}(\boldsymbol{r})$ solves the inhomogeneous
differential equation
\begin{equation}
\left[-\mathrm{i}c\hbar\boldsymbol{\alpha}\cdot\boldsymbol{\nabla}
+\beta mc^{2}-\frac{Ze^{2}}{(4\pi\epsilon_{0})r}-E^{(0)}\right]
\Psi_{\mu}^{(1)}(\boldsymbol{r})
=-\left[\frac{1}{2}ec\boldsymbol{\alpha}
\cdot(\boldsymbol{B}\times\boldsymbol{r})-E_{\mu}^{(1)}\right]
\Psi_{\mu}^{(0)}(\boldsymbol{r})
\label{2.10}
\end{equation}
($\boldsymbol{\alpha}$ and $\beta$ are the standard Dirac matrices),
subject to the usual regularity conditions and the orthogonality
constraint
\begin{equation}
\int_{\mathbb{R}^{3}}\mathrm{d}^{3}\boldsymbol{r}\:
\Psi_{\mu}^{(0)\dag}(\boldsymbol{r})
\Psi_{\mu'}^{(1)}(\boldsymbol{r})=0
\qquad ({\textstyle\mu,\mu'=\pm\frac{1}{2}}).
\label{2.11}
\end{equation}
The integral representation of $\Psi_{\mu}^{(1)}(\boldsymbol{r})$ is
\begin{equation}
\Psi_{\mu}^{(1)}(\boldsymbol{r})=-\frac{1}{2}ec\boldsymbol{B}
\cdot\int_{\mathbb{R}^{3}}\mathrm{d}^{3}\boldsymbol{r}'\:
\bar{G}\mbox{}^{(0)}(\boldsymbol{r},\boldsymbol{r}')
(\boldsymbol{r}'\times\boldsymbol{\alpha})
\Psi_{\mu}^{(0)}(\boldsymbol{r}'),
\label{2.12}
\end{equation}
where $\bar{G}\mbox{}^{(0)}(\boldsymbol{r},\boldsymbol{r}')$ is the
generalized Dirac--Coulomb Green function associated with the
ground-state energy level (\ref{2.1}).
%
%
\section{Analysis of electric multipole moments of the atom in the
magnetic field}
\label{III}
\setcounter{equation}{0}
After these preparatory steps, we set the problem: which electric
multipole moments, apart from the monopole one, characterize the
electronic cloud of the atom in the perturbed state
$\Psi_{\mu}(\boldsymbol{r})$? If $\rho_{\mu}(\boldsymbol{r})$ is the
electronic charge density for that state, the spherical components of
the $L$-th order electric multipole moment tensor are defined as
\begin{equation}
\mathcal{Q}_{LM\mu}=\sqrt{\frac{4\pi}{2L+1}}
\int_{\mathbb{R}^{3}}\mathrm{d}^{3}\boldsymbol{r}\:
r^{L}Y_{LM}^{*}(\boldsymbol{n}_{r})\rho_{\mu}(\boldsymbol{r}),
\label{3.1}
\end{equation}
where the asterisk denotes the complex conjugation and
$Y_{LM}(\boldsymbol{n}_{r})$ is the normalized spherical harmonic
defined according to the Condon--Shortley phase convention
\cite[chapter 5]{Vars75}. For the atom being in the state 
$\Psi_{\mu}(\boldsymbol{r})$, the density
$\rho_{\mu}(\boldsymbol{r})$ is given by
\begin{equation}
\rho_{\mu}(\boldsymbol{r})=\frac{-e\Psi_{\mu}^{\dag}(\boldsymbol{r})
\Psi_{\mu}(\boldsymbol{r})}{\int_{\mathbb{R}^{3}}
\mathrm{d}^{3}\boldsymbol{r}'\:\Psi_{\mu}^{\dag}(\boldsymbol{r}')
\Psi_{\mu}(\boldsymbol{r}')}.
\label{3.2}
\end{equation}
Using Eqs.\ (\ref{2.9}) and (\ref{2.11}), to the first order in the
perturbing field, one has
\begin{equation}
\rho_{\mu}(\boldsymbol{r})\simeq\rho_{\mu}^{(0)}(\boldsymbol{r})
+\rho_{\mu}^{(1)}(\boldsymbol{r}),
\label{3.3}
\end{equation}
with
\begin{equation}
\rho_{\mu}^{(0)}(\boldsymbol{r})
=-e\Psi_{\mu}^{(0)\dag}(\boldsymbol{r})
\Psi_{\mu}^{(0)}(\boldsymbol{r})
\label{3.4}
\end{equation}
and
\begin{equation}
\rho_{\mu}^{(1)}(\boldsymbol{r})
=-e\left[\Psi_{\mu}^{(1)\dag}(\boldsymbol{r})
\Psi_{\mu}^{(0)}(\boldsymbol{r})+\Psi_{\mu}^{(0)\dag}(\boldsymbol{r})
\Psi_{\mu}^{(1)}(\boldsymbol{r})\right].
\label{3.5}
\end{equation}
Accordingly, it follows that
\begin{equation}
\mathcal{Q}_{LM\mu}\simeq\mathcal{Q}_{LM\mu}^{(0)}
+\mathcal{Q}_{LM\mu}^{(1)},
\label{3.6}
\end{equation}
where
\begin{equation}
\mathcal{Q}_{LM\mu}^{(0)}=-e\sqrt{\frac{4\pi}{2L+1}}
\int_{\mathbb{R}^{3}}\mathrm{d}^{3}\boldsymbol{r}\:
\Psi_{\mu}^{(0)\dag}(\boldsymbol{r})
r^{L}Y_{LM}^{*}(\boldsymbol{n}_{r})\Psi_{\mu}^{(0)}(\boldsymbol{r})
\label{3.7}
\end{equation}
is the multipole moment for the unperturbed atom, whereas
\begin{equation}
\mathcal{Q}_{LM\mu}^{(1)}=\widetilde{\mathcal{Q}}_{LM\mu}^{(1)}
+(-)^{M}\widetilde{\mathcal{Q}}_{L,-M\mu}^{(1)*},
\label{3.8}
\end{equation}
with
\begin{equation}
\widetilde{\mathcal{Q}}_{LM\mu}^{(1)}=-e\sqrt{\frac{4\pi}{2L+1}}
\int_{\mathbb{R}^{3}}\mathrm{d}^{3}\boldsymbol{r}\:
\Psi_{\mu}^{(0)\dag}(\boldsymbol{r})
r^{L}Y_{LM}^{*}(\boldsymbol{n}_{r})\Psi_{\mu}^{(1)}(\boldsymbol{r}),
\label{3.9}
\end{equation}
is the first-order correction induced by the perturbing magnetic
field. To arrive at Eq.\ (\ref{3.8}), we have exploited the
well-known identity
\begin{equation}
Y_{LM}(\boldsymbol{n}_{r})=(-)^{M}Y_{L,-M}^{*}(\boldsymbol{n}_{r}).
\label{3.10}
\end{equation}
If the representation (\ref{2.12}) of 
$\Psi_{\mu}^{(1)}(\boldsymbol{r})$ is plugged into Eq.\ (\ref{3.9}),
this gives $\widetilde{\mathcal{Q}}_{LM\mu}^{(1)}$ in the form of
the double integral over $\mathbb{R}^{3}$:
\begin{equation}
\widetilde{\mathcal{Q}}_{LM\mu}^{(1)}
=\frac{1}{2}\sqrt{\frac{4\pi}{2L+1}}\,
e^{2}cB\int_{\mathbb{R}^{3}}\mathrm{d}^{3}\boldsymbol{r}
\int_{\mathbb{R}^{3}}\mathrm{d}^{3}\boldsymbol{r}'\:
\Psi_{\mu}^{(0)\dag}(\boldsymbol{r})
r^{L}Y_{LM}^{*}(\boldsymbol{n}_{r})
\bar{G}\mbox{}^{(0)}(\boldsymbol{r},\boldsymbol{r}')
\boldsymbol{n}_{z}\cdot(\boldsymbol{r}'\times\boldsymbol{\alpha})
\Psi_{\mu}^{(0)}(\boldsymbol{r}').
\label{3.11}
\end{equation}

Using Eq.\ (\ref{2.4}) and the explicit representations of the
spherical spinors $\Omega_{\mp1\mu}(\boldsymbol{n}_{r})$, it is easy
to show that the electronic charge density in the unperturbed atom,
$\rho_{\mu}^{(0)}(\boldsymbol{r})$, is spherically symmetric and
the integral in Eq.\ (\ref{3.7}) differs from zero only if $L=0$ and
$M=0$. Hence, for the ground state of the unperturbed atom it holds
that
\begin{equation}
\mathcal{Q}_{LM\mu}^{(0)}=-e\delta_{L0}\delta_{M0},
\label{3.12}
\end{equation}
i.e., all permanent electric multipole moments of the electronic
cloud other than the monopole one do vanish.

Next, we proceed to the analysis of the first-order induced multipole
moments $\mathcal{Q}_{LM\mu}^{(1)}$. To this end, we have to invoke
the multipole representation of the generalized Green function, which
is
\begin{eqnarray}
\bar{G}\mbox{}^{(0)}(\boldsymbol{r},\boldsymbol{r}')
&=& \frac{4\pi\epsilon_{0}}{e^{2}}
\sum_{\substack{\kappa=-\infty \\ (\kappa\neq0)}}^{\infty}
\sum_{m=-|\kappa|+1/2}^{|\kappa|-1/2}\frac{1}{rr'}
\nonumber \\
&& \hspace*{-5em}
\times\left(
\begin{array}{cc}
\bar{g}\mbox{}^{(0)}_{\kappa,(++)}(r,r')
\Omega_{\kappa m}(\boldsymbol{n}_{r})
\Omega_{\kappa m}^{\dag}(\boldsymbol{n}_{r}^{\prime}) &
-\mathrm{i}\bar{g}\mbox{}^{(0)}_{\kappa,(+-)}(r,r')
\Omega_{\kappa m}(\boldsymbol{n}_{r})
\Omega_{-\kappa m}^{\dag}(\boldsymbol{n}_{r}^{\prime}) \\
\mathrm{i}\bar{g}\mbox{}^{(0)}_{\kappa,(-+)}(r,r')
\Omega_{-\kappa m}(\boldsymbol{n}_{r})
\Omega_{\kappa m}^{\dag}(\boldsymbol{n}_{r}^{\prime}) &
\bar{g}\mbox{}^{(0)}_{\kappa,(--)}(r,r')
\Omega_{-\kappa m}(\boldsymbol{n}_{r})
\Omega_{-\kappa m}^{\dag}(\boldsymbol{n}_{r}^{\prime})
\end{array}
\right).
\nonumber \\
&&
\label{3.13}
\end{eqnarray}
If Eqs.\ (\ref{2.4}) and (\ref{3.13}) are inserted into Eq.\
(\ref{3.11}), the angular integration in the resulting formula may be
conveniently carried out with the aid of the identity \cite[Eq.\
(3.1.6)]{Szmy07}
\begin{eqnarray}
\boldsymbol{n}_{z}
\cdot(\boldsymbol{n}_{r}\times\boldsymbol{\sigma})\,
\Omega_{\kappa\mu}(\boldsymbol{n}_{r})
&=& \mathrm{i}\frac{4\mu\kappa}{4\kappa^{2}-1}
\Omega_{-\kappa\mu}(\boldsymbol{n}_{r})
+\mathrm{i}\frac{\sqrt{(\kappa+\frac{1}{2})^{2}-\mu^{2}}}{|2\kappa+1|}
\Omega_{\kappa+1,\mu}(\boldsymbol{n}_{r})
\nonumber \\
&& -\,\mathrm{i}
\frac{\sqrt{(\kappa-\frac{1}{2})^{2}-\mu^{2}}}{|2\kappa-1|}
\Omega_{\kappa-1,\mu}(\boldsymbol{n}_{r})
\label{3.14}
\end{eqnarray}
and the relation
\begin{eqnarray}
&& \hspace*{-3em}
\sqrt{\frac{4\pi}{2L+1}}
\oint_{4\pi}\mathrm{d}^{2}\boldsymbol{n}_{r}\:
\Omega_{\kappa\mu}^{\dag}(\boldsymbol{n}_{r})
Y_{LM}^{*}(\boldsymbol{n}_{r})
\Omega_{\kappa'\mu'}(\boldsymbol{n}_{r})
\nonumber \\
&& =\,(-)^{\mu'+1/2}2\sqrt{|\kappa\kappa'|}
\left(
\begin{array}{ccc}
|\kappa|-\frac{1}{2} & L & |\kappa'|-\frac{1}{2} \\
-\frac{1}{2} & 0 & \frac{1}{2}
\end{array}
\right)
\left(
\begin{array}{ccc}
|\kappa|-\frac{1}{2} & L & |\kappa'|-\frac{1}{2} \\
\mu & M & -\mu'
\end{array}
\right)
\Pi(l,L,l'),
\nonumber \\
&&
\label{3.15}
\end{eqnarray}
with
\begin{equation}
\Pi(l,L,l')
=\left\{
\begin{array}{ll}
1 & \textrm{for $l+L+l'$ even} \\
0 & \textrm{for $l+L+l'$ odd}.
\end{array}
\right.
\label{3.16}
\end{equation}
In Eq.\ (\ref{3.15}), {\footnotesize${\textstyle
\left(\begin{array}{ccc}j_{a} & j_{b} & j_{c} \\ 
m_{a} & m_{b} & m_{c}\end{array}\right)}$} denotes the Wigner's 3$j$
coefficient, whereas
\begin{equation}
l=\left|\kappa+\frac{1}{2}\right|-\frac{1}{2},
\label{3.17}
\end{equation} 
and similarly for $l'$. Exploiting the selection rules embodied in
Eq.\ (\ref{3.16}) and properties of the 3$j$ coefficients, one
deduces that the only case when
$\widetilde{\mathcal{Q}}_{LM\mu}^{(1)}$ does not vanish is the one
with $L=2$ and $M=0$. Since $\widetilde{\mathcal{Q}}_{20\mu}^{(1)}$
is real, from Eq.\ (\ref{3.8}) one has
\begin{equation}
\mathcal{Q}_{LM\mu}^{(1)}
=\mathcal{Q}_{20\mu}^{(1)}\delta_{L2}\delta_{M0},
\label{3.18}
\end{equation}
with $\mathcal{Q}_{20\mu}^{(1)}$ being given in the form of the
following double radial integral:
\begin{equation}
\mathcal{Q}_{20\mu}^{(1)}=\sgn(\mu)\frac{2}{15}(4\pi\epsilon_{0})cB
\int_{0}^{\infty}\mathrm{d}r\int_{0}^{\infty}\mathrm{d}r'\:
\left(
\begin{array}{cc}
P^{(0)}(r) & Q^{(0)}(r)
\end{array}
\right)
r^{2}\bar{\mathsf{G}}\mbox{}^{(0)}_{2}(r,r')r'
\left(
\begin{array}{c}
Q^{(0)}(r') \\
P^{(0)}(r')
\end{array}
\right),
\label{3.19}
\end{equation}
where
\begin{equation}
\bar{\mathsf{G}}\mbox{}^{(0)}_{\kappa}(r,r')
=\left(
\begin{array}{cc}
\bar{g}\mbox{}^{(0)}_{\kappa,(++)}(r,r') &
\bar{g}\mbox{}^{(0)}_{\kappa,(+-)}(r,r') \\
\bar{g}\mbox{}^{(0)}_{\kappa,(-+)}(r,r') &
\bar{g}\mbox{}^{(0)}_{\kappa,(--)}(r,r')
\end{array}
\right)
\label{3.20}
\end{equation}
is the radial generalized Green function.
%
%
\section{Evaluation of the induced electric quadrupole moment}
\label{IV} 
\setcounter{equation}{0}
In the preceding section, we have proved that, to the first order in
$B$, the only electric multipole moment induced in the electronic
cloud of the atom by the weak, static, uniform magnetic field is the
quadrupole ($L=2$) one, and that its only non-vanishing spherical
component is the one with $M=0$. Now, the Cartesian components of the
traceless tensor of the induced quadrupole moment,
\begin{equation}
\mathcal{Q}_{ij\mu}^{(1)}
=\int_{\mathbb{R}^{3}}\mathrm{d}^{3}\boldsymbol{r}\:
\frac{1}{2}\left(3r_{i}r_{j}-r^{2}\delta_{ij}\right)
\rho_{\mu}^{(1)}(\boldsymbol{r})
\qquad (i,j\in\{x,y,z\}),
\label{4.1}
\end{equation}
are related to its spherical components through
{\small
\begin{eqnarray}
\left(
\begin{array}{ccc}
\mathcal{Q}_{xx\mu}^{(1)} & \mathcal{Q}_{xy\mu}^{(1)} & 
\mathcal{Q}_{xz\mu}^{(1)} \\
\mathcal{Q}_{yx\mu}^{(1)} & \mathcal{Q}_{yy\mu}^{(1)} & 
\mathcal{Q}_{yz\mu}^{(1)} \\
\mathcal{Q}_{zx\mu}^{(1)} & \mathcal{Q}_{zy\mu}^{(1)} & 
\mathcal{Q}_{zz\mu}^{(1)}
\end{array}
\right)
\nonumber \\
&& \hspace*{-12em}
=\left(
\begin{array}{ccc}
-\frac{1}{2}\mathcal{Q}_{20\mu}^{(1)}+\sqrt{\frac{3}{8}}\,
[\mathcal{Q}_{22\mu}^{(1)}+\mathcal{Q}_{2-2\mu}^{(1)}] & 
\mathrm{i}\sqrt{\frac{3}{8}}\,
[\mathcal{Q}_{22\mu}^{(1)}-\mathcal{Q}_{2-2\mu}^{(1)}] & 
-\sqrt{\frac{3}{8}}\,
[\mathcal{Q}_{21\mu}^{(1)}-\mathcal{Q}_{2-1\mu}^{(1)}] \\
\mathrm{i}\sqrt{\frac{3}{8}}\,
[\mathcal{Q}_{22\mu}^{(1)}-\mathcal{Q}_{2-2\mu}^{(1)}] & 
-\frac{1}{2}\mathcal{Q}_{20\mu}^{(1)}-\sqrt{\frac{3}{8}}\,
[\mathcal{Q}_{22\mu}^{(1)}+\mathcal{Q}_{2-2\mu}^{(1)}] & 
-\mathrm{i}\sqrt{\frac{3}{8}}\,
[\mathcal{Q}_{21\mu}^{(1)}+\mathcal{Q}_{2-1\mu}^{(1)}] \\
-\sqrt{\frac{3}{8}}\,
[\mathcal{Q}_{21\mu}^{(1)}-\mathcal{Q}_{2-1\mu}^{(1)}] &
-\mathrm{i}\sqrt{\frac{3}{8}}\,
[\mathcal{Q}_{21\mu}^{(1)}+\mathcal{Q}_{2-1\mu}^{(1)}] &
\mathcal{Q}_{20\mu}^{(1)}
\end{array}
\right).
\nonumber \\
&&
\label{4.2}
\end{eqnarray}
}
\vspace*{-3ex}
\newline
Hence, it follows that in the Cartesian basis all the off-diagonal
elements of the quadrupole moment induced in the system under study
do vanish:
\begin{equation}
\mathcal{Q}_{ij\mu}^{(1)}=0
\qquad (\textrm{$i\neq j$; $i,j\in\{x,y,z\}$}),
\label{4.3}
\end{equation}
while the diagonal elements are given by
\begin{equation}
\mathcal{Q}_{xx\mu}^{(1)}=\mathcal{Q}_{yy\mu}^{(1)}
=-\frac{1}{2}\mathcal{Q}_{zz\mu}^{(1)}
=-\frac{1}{2}\mathcal{Q}_{20\mu}^{(1)}.
\label{4.4}
\end{equation}
Such a structure of the Cartesian representation of the quadrupole
moment tensor is characteristic for systems possessing the rotational
symmetry around the $z$ axis (recall that in our case the $z$ axis is
directed along the magnetic field).

It remains to evaluate the double integral in Eq.\ (\ref{3.19}). For
this purpose, we shall make use of the Sturmian expansion of the
pertinent generalized radial Dirac--Coulomb Green function for the
atomic ground state, which is \cite{Szmy97}
\begin{equation}
\bar{\mathsf{G}}\mbox{}^{(0)}_{\kappa}(r,r')
=\sum_{n_{r}=-\infty}^{\infty}
\frac{1}{\mu_{n_{r}\kappa}^{(0)}-1}
\left(
\begin{array}{c}
S_{n_{r}\kappa}^{(0)}(r) \\
T_{n_{r}\kappa}^{(0)}(r)
\end{array}
\right)
\left(
\begin{array}{cc}
\mu_{n_{r}\kappa}^{(0)}S_{n_{r}\kappa}^{(0)}(r') 
& T_{n_{r}\kappa}^{(0)}(r')
\end{array}
\right)
\qquad (\kappa\neq-1).
\label{4.5}
\end{equation}
Here
\begin{subequations}
\begin{eqnarray}
S_{n_{r}\kappa}^{(0)}(r)
&=& \sqrt{\frac{(1+\gamma_{1})(|n_{r}|+2\gamma_{\kappa})|n_{r}|!}
{2ZN_{n_{r}\kappa}(N_{n_{r}\kappa}-\kappa)
\Gamma(|n_{r}|+2\gamma_{\kappa})}}
\nonumber \\
&& \times\left(\frac{2Zr}{a_{0}}\right)^{\gamma_{\kappa}}
\textrm{e}^{-Zr/a_{0}}
\left[L_{|n_{r}|-1}^{(2\gamma_{\kappa})}\left(\frac{2Zr}{a_{0}}\right)
+\frac{\kappa-N_{n_{r}\kappa}}{|n_{r}|+2\gamma_{\kappa}}
L_{|n_{r}|}^{(2\gamma_{\kappa})}\left(\frac{2Zr}{a_{0}}\right)\right]
\label{4.6a}
\end{eqnarray}
and
\begin{eqnarray}
T_{n_{r}\kappa}^{(0)}(r)
&=& \sqrt{\frac{(1-\gamma_{1})(|n_{r}|+2\gamma_{\kappa})|n_{r}|!}
{2ZN_{n_{r}\kappa}(N_{n_{r}\kappa}-\kappa)
\Gamma(|n_{r}|+2\gamma_{\kappa})}}
\nonumber \\
&& \times\left(\frac{2Zr}{a_{0}}\right)^{\gamma_{\kappa}}
\textrm{e}^{-Zr/a_{0}}
\left[L_{|n_{r}|-1}^{(2\gamma_{\kappa})}
\left(\frac{2Zr}{a_{0}}\right)
-\frac{\kappa-N_{n_{r}\kappa}}{|n_{r}|+2\gamma_{\kappa}}
L_{|n_{r}|}^{(2\gamma_{\kappa})}\left(\frac{2Zr}{a_{0}}\right)\right]
\label{4.6b}
\end{eqnarray}
\label{4.6}
\end{subequations}
\newline
[with $L_{n}^{(\alpha)}(x)$ denoting the generalized Laguerre
polynomial \cite{Magn66}; we define $L_{-1}^{(\alpha)}(\rho)\equiv0$]
are the radial Dirac--Coulomb Sturmian functions associated with the
hydrogenic ground-state energy level, whereas
\begin{equation}
\mu_{n_{r}\kappa}^{(0)}
=\frac{|n_{r}|+\gamma_{\kappa}+N_{n_{r}\kappa}}{\gamma_{1}+1},
\label{4.7}
\end{equation}
with
\begin{equation}
N_{n_{r}\kappa}
=\pm\sqrt{(|n_{r}|+\gamma_{\kappa})^{2}+(\alpha Z)^{2}}
=\pm\sqrt{|n_{r}|^{2}+2|n_{r}|\gamma_{\kappa}+\kappa^{2}}
\label{4.8}
\end{equation}
being the `apparent principal quantum number' (notice that it may
assume positive as well as negative values!); the following sign
convention applies to the definition (\ref{4.8}): the plus sign
should be chosen for $n_{r}>0$ and the minus one for $n_{r}<0$; for
$n_{r}=0$ one chooses the plus sign if $\kappa<0$ and the minus sign
if $\kappa>0$. Insertion of the separable expansion (\ref{4.5}) into
the right-hand side of the formula in Eq.\ (\ref{3.19}) leads to the
following expression for $\mathcal{Q}_{20\mu}^{(1)}$:
\begin{eqnarray}
\mathcal{Q}_{20\mu}^{(1)} 
&=& \sgn(\mu)\frac{2}{15}(4\pi\epsilon_{0})cB
\sum_{n_{r}=-\infty}^{\infty}\frac{1}{\mu_{n_{r}2}^{(0)}-1}
\int_{0}^{\infty}\mathrm{d}r\:r^{2}
\left[P^{(0)}(r)S_{n_{r}2}^{(0)}(r)
+Q^{(0)}(r)T_{n_{r}2}^{(0)}(r)\right]
\nonumber \\
&& \times\int_{0}^{\infty}\mathrm{d}r'\:r'
\left[\mu_{n_{r}2}^{(0)}Q^{(0)}(r')S_{n_{r}2}^{(0)}(r')
+P^{(0)}(r')T_{n_{r}2}^{(0)}(r')\right].
\label{4.9}
\end{eqnarray}
The radial integrals in Eq.\ (\ref{4.9}) may be taken after one makes
use of Eq.\ (\ref{4.7}) and of the explicit representations of the
radial functions $P^{(0)}(r)$, $Q^{(0)}(r)$ and the radial Sturmians
$S_{n_{r}2}^{(0)}(r)$, $T_{n_{r}2}^{(0)}(r)$, given by Eqs.\
(\ref{2.6}) and (\ref{4.6}), respectively. Exploiting the known
integral formula \cite[Eq.\ (7.414.11)]{Grad07}
\begin{equation}
\int_{0}^{\infty}\mathrm{d}x\:
x^{\gamma}\mathrm{e}^{-x}L_{n}^{(\alpha)}(x)
=\frac{\Gamma(\gamma+1)\Gamma(n+\alpha-\gamma)}
{n!\Gamma(\alpha-\gamma)}
\qquad [\Real(\gamma)>-1]
\label{4.10}
\end{equation}
and the trivial but extremely useful identity
\begin{equation}
\gamma_{2}^{2}=\gamma_{1}^{2}+3,
\label{4.11}
\end{equation}
one arrives at
\begin{eqnarray}
\mathcal{Q}_{20\mu}^{(1)} 
&=& -\,\sgn(\mu)\frac{\alpha^{2}ea_{0}^{2}}{Z^{4}}
\frac{B}{B_{0}}\frac{\Gamma^{2}(\gamma_{1}+\gamma_{2}+3)}
{480(4\gamma_{1}+1)\Gamma(2\gamma_{1}+1)
\Gamma^{2}(\gamma_{2}-\gamma_{1}-2)}
\nonumber \\
&& \times\sum_{n_{r}=-\infty}^{\infty}
\frac{\Gamma(|n_{r}|+\gamma_{2}-\gamma_{1}-3)
\Gamma(|n_{r}|+\gamma_{2}-\gamma_{1}-2)}
{|n_{r}|!\Gamma(|n_{r}|+2\gamma_{2}+1)}
\frac{N_{n_{r}2}-2}{N_{n_{r}2}}
\nonumber \\
&& \times
\frac{(|n_{r}|+\gamma_{2}-3\gamma_{1}-3-\gamma_{1}N_{n_{r}2})
(3|n_{r}|+\gamma_{1}+3\gamma_{2}+1+3N_{n_{r}2})}
{|n_{r}|+\gamma_{2}-\gamma_{1}-1+N_{n_{r}2}}.
\label{4.12}
\end{eqnarray}
The above result may be simplified considerably if in the series
$\sum_{n_{r}=-\infty}^{\infty}(\cdots)$ one collects together terms
with the same absolute value of the summation index $n_{r}$ (the
Sturmian radial quantum number). Proceeding in that way, after much
labor, using Eq.\ (\ref{4.8}) and again the identity (\ref{4.11}),
one finds that
\begin{eqnarray}
\mathcal{Q}_{20\mu}^{(1)} 
&=& \sgn(\mu)\frac{\alpha^{2}ea_{0}^{2}}{Z^{4}}
\frac{B}{B_{0}}\frac{\Gamma^{2}(\gamma_{1}+\gamma_{2}+3)}
{240(4\gamma_{1}+1)\Gamma(2\gamma_{1}+1)
\Gamma^{2}(\gamma_{2}-\gamma_{1}-2)}
\nonumber \\
&& \times\sum_{n_{r}=0}^{\infty}
\frac{\Gamma(n_{r}+\gamma_{2}-\gamma_{1}-3)
\Gamma(n_{r}+\gamma_{2}-\gamma_{1}-2)}
{n_{r}!(n_{r}+\gamma_{2}-\gamma_{1})\Gamma(n_{r}+2\gamma_{2}+1)}
\nonumber \\
&& \times[(2\gamma_{1}-1)(n_{r}+\gamma_{2}-\gamma_{1}-3)
(n_{r}+\gamma_{2}-\gamma_{1})
+2(\gamma_{1}+1)(2\gamma_{1}+1)
(n_{r}+\gamma_{2}-\gamma_{1})
\nonumber \\
&& \qquad -\,6(\gamma_{1}+1)].
\label{4.13}
\end{eqnarray}
It is possible to achieve a further simplification. To this end, we
express the right-hand side of Eq.\ (\ref{4.13}) in terms of the
hypergeometric functions $_{2}F_{1}$ and $_{3}F_{2}$ of the unit
argument. Since it holds that
\begin{equation}
\sum_{n=0}^{\infty}\frac{\Gamma(n+a_{1})\Gamma(n+a_{2})}
{n!\Gamma(n+b)}
=\frac{\Gamma(a_{1})\Gamma(a_{2})}{\Gamma(b)}\,
{}_{2}F_{1}
\left(
\begin{array}{c}
a_{1},a_{2} \\
b
\end{array}
;1
\right)
\label{4.14}
\end{equation}
and
\begin{equation}
\sum_{n=0}^{\infty}
\frac{\Gamma(n+a_{1})\Gamma(n+a_{2})}{n!(n+a_{3})\Gamma(n+b)}
=\frac{\Gamma(a_{1})\Gamma(a_{2})}{a_{3}\Gamma(b)}\,
{}_{3}F_{2}
\left(
\begin{array}{c}
a_{1},a_{2},a_{3} \\
a_{3}+1,b
\end{array}
;1
\right),
\label{4.15}
\end{equation}
Eq.\ (\ref{4.13}) becomes
\begin{eqnarray}
\mathcal{Q}_{20\mu}^{(1)} 
&=& -\,\sgn(\mu)\frac{\alpha^{2}ea_{0}^{2}}{Z^{4}}
\frac{B}{B_{0}}\frac{\Gamma(\gamma_{1}+\gamma_{2}+3)
\Gamma(\gamma_{1}+\gamma_{2}+4)}
{1440(\gamma_{1}+1)(4\gamma_{1}+1)\Gamma(2\gamma_{1}+1)
\Gamma(2\gamma_{2}+1)}
\nonumber \\
&& \times\left[
(2\gamma_{1}-1)(\gamma_{2}-\gamma_{1}-3)\,
{}_{2}F_{1}\left(
\begin{array}{c}
\gamma_{2}-\gamma_{1}-2,\gamma_{2}-\gamma_{1}-2 \\
2\gamma_{2}+1
\end{array}
;1
\right)
\right.
\nonumber \\
&& \qquad +\,2(\gamma_{1}+1)(2\gamma_{1}+1)\,
{}_{2}F_{1}\left(
\begin{array}{c}
\gamma_{2}-\gamma_{1}-3,\gamma_{2}-\gamma_{1}-2 \\
2\gamma_{2}+1
\end{array}
;1
\right)
\nonumber \\
&& \qquad -\left.
\frac{6(\gamma_{1}+1)}{\gamma_{2}-\gamma_{1}}\,
{}_{3}F_{2}\left(
\begin{array}{c}
\gamma_{2}-\gamma_{1}-3,\gamma_{2}-\gamma_{1}-2,
\gamma_{2}-\gamma_{1} \\
\gamma_{2}-\gamma_{1}+1,2\gamma_{2}+1
\end{array}
;1
\right)
\right].
\label{4.16}
\end{eqnarray}
The two $_{2}F_{1}$ functions may be then eliminated with the aid of
the Gauss' identity \cite[Eq.\ (9.122.1)]{Grad07}
\begin{equation}
{}_{2}F_{1}
\left(
\begin{array}{c}
a_{1},a_{2} \\
b
\end{array}
;1
\right)
=\frac{\Gamma(b)\Gamma(b-a_{1}-a_{2})}
{\Gamma(b-a_{1})\Gamma(b-a_{2})}
\qquad [\Real(b-a_{1}-a_{2})>0].
\label{4.17}
\end{equation}
After some simple algebra, we obtain
\begin{eqnarray}
\mathcal{Q}_{20\mu}^{(1)} 
&=& \sgn(\mu)\frac{\alpha^{2}ea_{0}^{2}}{Z^{4}}
\frac{B}{B_{0}}\frac{\Gamma(2\gamma_{1}+5)}
{720(4\gamma_{1}+1)\Gamma(2\gamma_{1}+1)}
\bigg[-2(2\gamma_{1}^{2}+3\gamma_{1}+4)
\nonumber \\
&& +\,\frac{(\gamma_{1}+\gamma_{2})\Gamma(\gamma_{1}+\gamma_{2}+3)
\Gamma(\gamma_{1}+\gamma_{2}+4)}
{\Gamma(2\gamma_{1}+5)\Gamma(2\gamma_{2}+1)}\,
{}_{3}F_{2}
\left(
\begin{array}{c}
\gamma_{2}-\gamma_{1}-3,\gamma_{2}-\gamma_{1}-2,
\gamma_{2}-\gamma_{1} \\
\gamma_{2}-\gamma_{1}+1,2\gamma_{2}+1
\end{array}
;1
\right)
\bigg].
\nonumber \\
&&
\label{4.18}
\end{eqnarray}
Various other equivalent representations of
$\mathcal{Q}_{20\mu}^{(1)}$ may be derived from Eq.\ (\ref{4.18})
with the help of recurrence relations obeyed by the $_{3}F_{2}(1)$
function. For instance, if one uses repeatedly the relation
\begin{eqnarray}
&& {}_{3}F_{2}
\left(
\begin{array}{c}
a_{1},a_{2},a_{3} \\
a_{3}+1,b
\end{array}
;1
\right)
=-\frac{a_{3}}{a_{1}-a_{3}}
\frac{\Gamma(b)\Gamma(b-a_{1}-a_{2})}
{\Gamma(b-a_{1})\Gamma(b-a_{2})}
+\frac{a_{1}}{a_{1}-a_{3}}\,
{}_{3}F_{2}
\left(
\begin{array}{c}
a_{1}+1,a_{2},a_{3} \\
a_{3}+1,b
\end{array}
;1
\right)
\nonumber \\
&& \hspace*{20em}
[\Real(b-a_{1}-a_{2})>0],
\label{4.19}
\end{eqnarray}
and its analogue with $a_{1}$ and $a_{2}$ interchanged, Eq.\
(\ref{4.18}) is casted into the slightly more compact expression
\begin{eqnarray}
\mathcal{Q}_{20\mu}^{(1)} 
&=& \sgn(\mu)\frac{\alpha^{2}ea_{0}^{2}}{Z^{4}}
\frac{B}{B_{0}}\frac{\Gamma(2\gamma_{1}+5)}{1440\,\Gamma(2\gamma_{1})}
\left[-1+\frac{(\gamma_{1}+1)(\gamma_{1}+\gamma_{2})
\Gamma(\gamma_{1}+\gamma_{2}+2)\Gamma(\gamma_{1}+\gamma_{2}+3)}
{\gamma_{1}\Gamma(2\gamma_{1}+5)\Gamma(2\gamma_{2}+1)}\right.
\nonumber \\
&& \qquad \times\left.
{}_{3}F_{2}
\left(
\begin{array}{c}
\gamma_{2}-\gamma_{1}-2,\gamma_{2}-\gamma_{1}-1,
\gamma_{2}-\gamma_{1} \\
\gamma_{2}-\gamma_{1}+1,2\gamma_{2}+1
\end{array}
;1
\right)
\right].
\label{4.20}
\end{eqnarray}

Before concluding, it seems worthwhile to investigate the
approximation to Eq.\ (\ref{4.20}) for \mbox{$\alpha Z\ll1$}. Then
one has
\begin{equation}
\gamma_{\kappa}\simeq|\kappa|-\frac{(\alpha Z)^{2}}{2|\kappa|}
\qquad (\kappa\in\mathbb{Z}\setminus\{0\}),
\label{4.21}
\end{equation}
and consequently
\begin{eqnarray}
\Gamma(\gamma_{\kappa}+\gamma_{\kappa'}+k)
&\simeq& (|\kappa|+|\kappa'|+k-1)!
\left[1-\frac{1}{2}\left(\frac{1}{|\kappa|}+\frac{1}{|\kappa'|}\right)
\psi\left(|\kappa|+|\kappa'|+k\right)(\alpha Z)^{2}\right]
\nonumber \\
&& \hspace*{15em} 
(\textrm{$\kappa,\kappa'\in\mathbb{Z}\setminus\{0\}$,
$k\in\mathbb{N}$}),
\label{4.22}
\end{eqnarray}
where
\begin{equation}
\psi(\zeta)=\frac{1}{\Gamma(\zeta)}
\frac{\mathrm{d}\Gamma(\zeta)}{\mathrm{d}\zeta}
\label{4.23}
\end{equation}
is the digamma function. Using Eq.\ (\ref{4.22}) and the recurrence
relation
\begin{equation}
\psi(\zeta+1)=\psi(\zeta)+\frac{1}{\zeta}
\label{4.24}
\end{equation}
yields
\begin{equation}
\frac{(\gamma_{1}+1)(\gamma_{1}+\gamma_{2})
\Gamma(\gamma_{1}+\gamma_{2}+2)\Gamma(\gamma_{1}+\gamma_{2}+3)}
{\gamma_{1}\Gamma(2\gamma_{1}+5)\Gamma(2\gamma_{2}+1)}
\simeq1+\frac{13}{60}(\alpha Z)^{2}.
\label{4.25}
\end{equation}
Since at the same time, to the second order in $\alpha Z$, it holds
that
\begin{equation}
{}_{3}F_{2}
\left(
\begin{array}{c}
\gamma_{2}-\gamma_{1}-2,\gamma_{2}-\gamma_{1}-1,
\gamma_{2}-\gamma_{1} \\
\gamma_{2}-\gamma_{1}+1,2\gamma_{2}+1
\end{array}
;1
\right)
\simeq1-\frac{(\alpha Z)^{2}}{40},
\label{4.26}
\end{equation}
the sought approximation to $\mathcal{Q}_{20\mu}^{(1)}$ is
\begin{equation}
\mathcal{Q}_{20\mu}^{(1)}\simeq\sgn(\mu)
\frac{23}{240}\frac{\alpha^{4}ea_{0}^{2}}{Z^{2}}\frac{B}{B_{0}}
\qquad (\alpha Z\ll1).
\label{4.27}
\end{equation}
It is seen from Eqs.\ (\ref{4.27}) and (\ref{1.1}) that for the
hydrogen atom ($Z=1)$ and for the perturbing magnetic field
comparable with the intra-atomic magnetic field, i.e., for $B\simeq
B_{0}$, the first-order quadrupole moment $\mathcal{Q}_{20\mu}^{(1)}$
predicted by the relativistic formalism is of the same order of
magnitude as the second-order moment $\mathcal{Q}_{20}^{(2)}$
obtained from the non-relativistic theory.
%
%
\section{Conclusions}
\label{V}
\setcounter{equation}{0}
Earlier calculations of the magnetic field-induced electric
quadrupole moment in the ground state of the hydrogen-like atom,
based on the non-relativistic atomic model, predicted the quadratic
dependence of that moment on the field strength in the low-field
regime. In the present paper, we have shown that if the relativity is
taken into account and considerations are based on the Dirac rather
than the Schr{\"o}dinger or the Pauli equation for the electron, the
leading term in the expansion of the induced electric quadrupole
moment in powers of the field strength appears to be linear, not
quadratic. The calculations of the actual value of that moment we
have carried out in Sec.\ \ref{IV} provide a still another example of
the usefulness of the Sturmian expansion of the generalized
Dirac--Coulomb Green function \cite{Szmy97} for analytical
determination of electromagnetic properties of the relativistic
one-electron atom.
%
%

%

\begin{thebibliography}{99}
\bibitem{Coul56}
   C.\ A.\ Coulson, M.\ J.\ Stephen,
   Proc.\ Phys.\ Soc.\ A 69 (1956) 777
\bibitem{Turb87}
   A.\ V.\ Turbiner,
   Yad.\ Fiz.\ 46 (1987) 204
\bibitem{Pote01}
   A.\ Y.\ Potekhin, A.\ V.\ Turbiner,
   Phys.\ Rev.\ A 63 (2001) 065402
\bibitem{Szmy97}
   R.\ Szmytkowski,
   J.\ Phys.\ B 30 (1997) 825
   [erratum J.\ Phys.\ B 30 (1997) 2747; 
   addendum arXiv:physics/9902050]
\bibitem{Szmy02}
   R.\ Szmytkowski,
   J.\ Phys.\ B 35 (2002) 1379
\bibitem{Szmy04}
   R.\ Szmytkowski, K.\ Mielewczyk,
   J.\ Phys.\ B 37 (2004) 3961
\bibitem{Miel06}
   K.\ Mielewczyk, R.\ Szmytkowski,
   Phys.\ Rev.\ A 73 (2006) 022511
   [erratum Phys.\ Rev.\ A 73 (2006) 039908]
\bibitem{Szmy11}
   R.\ Szmytkowski, P.\ Stefa{\'n}ska,
   arXiv:1102.1811
\bibitem{Stef11}
   P.\ Stefa{\'n}ska, R.\ Szmytkowski,
   Int.\ J.\ Quantum Chem.\ doi:10.1002/qua.23108
   [preprint arXiv:1102.3853]
\bibitem{Szmy07}
   R.\ Szmytkowski,
   J.\ Math.\ Chem.\ 42 (2007) 397
   [preprint arXiv:1011.3433]
\bibitem{Vars75}
   D.\ A.\ Varshalovich, A.\ N.\ Moskalev, V.\ K.\ Khersonskii,
   Quantum Theory of Angular Momentum,
   Nauka, Leningrad, 1975 (in Russian)
\bibitem{Magn66}
   W.\ Magnus, F.\ Oberhettinger, R.\ P.\ Soni,
   Formulas and Theorems for the Special Functions of Mathematical
   Physics, 3rd ed.,
   Springer, Berlin, 1966
\bibitem{Grad07}
   I.\ S.\ Gradshteyn, I.\ M.\ Ryzhik,
   Table of Integrals, Series, and Products, 7th ed.,
   Elsevier, Amsterdam, 2007
\end{thebibliography}
\end{document}